\documentclass[twocolumn,pre,floatfix]{revtex4}
\usepackage{psfrag,epsfig,amsfonts,amssymb,amsmath,wasysym,bm}
\usepackage{dcolumn}
\usepackage{bbold}
\usepackage[normalem]{ulem}
\usepackage{color}

\newcommand{\id}{\mathbb 1}
\newcommand{\psip}{\psi^{\perp\!}}
\newcommand{\pmax}{p_{\rm max}}
\newcommand{\Prob}{{\rm Prob}}
\newcommand{\Aobs}{{\cal A}}

\newcommand{\hr}{{\cal H}}

\providecommand{\norm}[1]{\|#1\|}

\begin{document}

\title{Temporal relaxation of gapped many-body quantum systems}

\author{Peter Reimann$^{1}$, 
Ben N. Balz$^{1}$,
Jonas Richter$^{2}$, 
and Robin Steinigeweg$^{2}$}
\affiliation{$^{1}$Fakult\"at f\"ur Physik, 
Universit\"at Bielefeld, 
33615 Bielefeld, Germany
\\
$^{2}$Department of Physics, 
University of Osnabr\"uck, 
49069 Osnabr\"uck, Germany}

\begin{abstract}
Typicality of the orthogonal dynamics (TOD)
is established as 
a generic feature of
temporal relaxation processes in 
isolated many-body quantum systems.
The basic idea in the simplest case is that 
the transient non-equilibrium behavior is
mainly governed by the component of 
the time-evolved system state parallel
to the initial state, while the orthogonal
component
appears as equilibrated
right from the beginning.
The main emphasis is laid on the 
largely unexplored and particularly
challenging
case that
one energy 
level exhibits a much larger population than 
all the others.
Important examples are gapped many-body 
systems at low energies, for instance due 
to a quantum quench.
A general analytical prediction is derived and
is found to compare very well with various
numerically exact results.
\end{abstract}

\maketitle

\section{Introduction}
\label{s1}
Equilibration and thermalization
of possibly 
far-from-equilibrium initial states
in isolated many-body quantum systems have
recently attracted a considerable amount of
theoretical and experimental
interest \cite{gog16,dal16,tas16,bor16,lan16,mor18}.
In particular, despite the well-known 
quantum revival and reversibility properties 
of the system's unitary time evolution \cite{hob71}, 
it could be shown 
\cite{rei08,lin09,sho11,rei12,sho12,bal16}
that rather weak
assumptions 
are sufficient to guarantee
equilibration in the following sense:
The expectation values of a given 
observable stay extremely close to 
a constant value for the vast majority 
of all sufficiently
late times, i.e., after the initial transients 
have died out.
As detailed in 
\cite{rei08,lin09,sho11,rei12,sho12,bal16}, 
to demonstrate equilibration in this sense it is sufficient
that the system's energy
spectrum does not exhibit some
highly non-generic features,
and that every single of 
the extremely numerous energy 
levels is only weakly 
populated by the initial 
state.
Somewhat less appreciated so far 
is the fact that the latter condition 
is actually not an indispensable
prerequisite:
Namely, equilibration is still guaranteed
if there is one exceptional, non-small 
energy level population 
\cite{rei12,bal16}.
In particular, the requirement from 
\cite{lin09,sho11,sho12}
that the so-called effective 
dimension or inverse participation 
ration (IPR) must be large, 
is then no longer fulfilled.

The main goal of our present work
is to gain analytical insight into 
the detailed temporal decay of the 
above mentioned initial transients,
especially (but not exclusively) in
cases with a non-small population 
of one level.
The first reason is that their temporal 
relaxation behavior turns out to be 
particularly non-trivial to predict.
Second, they have been hardly considered 
before, yet seem of considerable conceptual 
and practical interest.

Typical examples 
where such a scenario may naturally arise 
are systems with an energy gap between 
the ground state and the excited states. 
Starting in the ground state
(or some more general low energy state)
and then suddenly changing a system 
parameter (quantum quench) may
often result in a far-from-equilibrium 
initial state with a large population 
of the ``new'' (post-quench)
ground state and a small population 
of all other states \cite{maz16}.
For instance, such gaps commonly arise in
solid-state insulators
as a consequence of their electronic band 
structure.
Moreover, other possibilities than
quenches to generate initial states
with one non-small level population
are easily conceivable.

Further examples 
to which our present
theory should be readily 
applicable are 
systems which, instead of a gap, exhibit
a highly degenerate ground state, such as
sawtooth Heisenberg chains, 
or kagom\'e, checkerboard,
and pyrochlore spin lattices,
see \cite{kri14} and references therein.
Finally, also more conventional situations
without any non-small populations, gaps, 
or high degeneracies will be included
as special cases.

\section{General Framework}
\label{s2}
We consider an isolated many-body quantum system, 
modeled by a Hamiltonian 
$H$ with eigenvalues $E_n$ and 
eigenvectors $|n\rangle$.
A given initial state $|\psi_0\rangle$ 
with components $c_n:=\langle n| \psi_0\rangle$
then evolves in time as
($\hbar =1$)
\begin{eqnarray}
|\psi_t\rangle=e^{-iHt}|\psi_0\rangle
=\sum_{n=0}^N c_n\, e^{-iE_nt}\, |n\rangle
\ ,
\label{1}
\end{eqnarray}
where, depending on the 
specific model under consideration, $N$ may be 
large but finite or infinite.
Further, $p_n:=|c_n|^2$ are the
level populations with
$\sum_{n=0}^Np_n=1$.

In case that, say, $E_n$ is degenerate,
we choose $|n\rangle$ as the (normalized) 
projection of $|\psi_0\rangle$ 
onto the corresponding eigenspace
$\hr_n$.
Hence, all other
basis vectors of $\hr_n$
play no role 
and can be omitted in (\ref{1}).
In other words, while the actual system
may well exhibit degeneracies,
we can and will exclude them in (\ref{1}).

Choosing $n=0$ for the specific
level with a possibly non-small population 
$p_0$,
and ignoring the trivial case $p_0=1$, 
we rewrite (\ref{1}) as
\begin{eqnarray}
|\psi_t\rangle
& = &
c_0\,
e^{-i E_0 t}
|0\rangle
+ \sqrt{1-p_0}\, |\psi'_t\rangle
\label{2}
\\
 |\psi'_t\rangle
 & := &
\sum_{n=1}^N c'_n\, e^{-iE_nt}\, |n\rangle\ ,\ c'_n:=\frac{c_n}{\sqrt{1-p_0}}
\ .
\label{3}
\end{eqnarray}
It follows that $ |\psi'_t\rangle$ and $|0\rangle$ are
orthonormalized.
Defining
\begin{eqnarray}
\alpha'_t
&:=&
\langle\psi'_0|\psi'_t\rangle\, ,\ 
P:=\id -
|\psi'_0\rangle\langle\psi'_0|
\ ,
\label{4}
\\
|\psip_t\rangle
& := &
P|\psi'_t\rangle/\beta'_t\, , \ 
\beta'_t := \langle\psi'_t|P|\psi'_t\rangle^{1/2}
\ ,
\label{5}
\end{eqnarray}
one readily concludes 
\cite{f1}
that
$|\psi'_t\rangle=\alpha'_t|\psi'_0\rangle+\beta'_t|\psip_t\rangle$
and thus
\begin{eqnarray}
|\psi_t\rangle
& = &
c_0\, 
e^{-i E_0 t}
|0\rangle
+ \alpha_t|\psi'_0\rangle+\beta_t|\psip_t\rangle
\ ,
\label{6}
\\
\alpha_t 
&:=&
\alpha'_t\sqrt{1-p_0}\, ,\ 
\beta_t :=\beta'_t\sqrt{1-p_0}
\ .
\label{7}
\end{eqnarray}
Furthermore, $|0\rangle$, $|\psi'_0\rangle$, and $|\psip_t\rangle$
are normalized and pairwise orthogonal, 
implying $\langle\psip_t|\psi_0\rangle=0$ and
\begin{eqnarray}
\beta_t=\sqrt{1-p_0-|\alpha_t|^2}
 \ .
\label{8}
\end{eqnarray}

The main virtue of (\ref{6}), which will be heavily
exploited in the following, is that the system
dynamics has been decomposed into three 
orthogonal components, two of them pointing
into the ``special directions'' $|0\rangle$ 
and $|\psi'_0\rangle$ (and encapsulating the 
initial state $|\psi_0\rangle$), 
and an ``orthogonal rest'' $|\psip_t\rangle$.

\section{Analytical prediction}
\label{s3}
Given an observable (Hermitian operator) $A$,
we can infer from (\ref{6}) and (\ref{8}) that
\begin{eqnarray}
\Aobs_t & := &
\langle\psi_t|A|\psi_t\rangle 
=
p_0 \langle 0|A|0\rangle
+|\alpha_t|^2\langle\psi'_0|A|\psi'_0\rangle
\nonumber
\\
& & +\ 
\beta_t^2\, \langle\psip_t|A|\psip_t\rangle 
+ q_t +q_t^\ast
\ ,\ \ 
\label{9}
\\
q_t & := & 
c_0^\ast
\, e^{iE_0t}
\big(
\alpha_t\langle0|A|\psi'_0\rangle+r_t
\big) 
+
\alpha_t^\ast s_t
\ ,
\label{10}
\\
r_t & := & \beta_t\langle 0|A |\psip_t\rangle 
\, , \ 
s_t := \beta_t\langle \psi'_0|A |\psip_t\rangle
\ .
\label{11}
\end{eqnarray}
Exploiting (\ref{5}) and (\ref{7}), we rewrite
$r_t$ from (\ref{11}) as
$\langle v|\psi'_t\rangle$,
where
$|v\rangle:=\sqrt{1-p_0}\, PA|0\rangle$.
With (\ref{3}) we thus obtain
\begin{eqnarray}
r_t = \sum_{n=1}^N b_n\, e^{-iE_nt}
\, ,\ 
b_n:=c_n'\langle v|n\rangle
\ .
\label{12}
\end{eqnarray}
Indicating time averages over all $t\geq 0$ by an overbar,
i.e., $\overline{\bullet} := \lim_{T\to \infty} \int_0^T \bullet \, 
\text{d}t/T$, and recalling that
degeneracies are excluded in (\ref{1}),
it follows that
\begin{eqnarray}
\overline{|r_t|^2} 
= \sum_{n=1}^N |b_n|^2 
= \sum_{n=1}^N |c'_n|^2 \langle v|n\rangle  \langle n|v\rangle 
\ .
\label{13}
\end{eqnarray}
Hence, $\overline{|r_t|^2}$ can be upper bounded
by $\langle v|v\rangle \max_{n\geq 1}|c'_n|^2$,
and with (\ref{3}) by
$\kappa\pmax$, where
$\kappa := \langle 0|APA|0\rangle$ and
$\pmax:= \max_{n\geq 1}p_n$.
Denoting by $\norm{\cdot}$ the operator norm, and 
exploiting that $\norm{P}=1$ since $P$ from (\ref{4})
is a projector, it follows that $\kappa\leq\norm{APA}\leq\norm{A}^2$,
and thus
$\overline{|r_t|^2} \leq \norm{A}^2\, \pmax$.
Considering $|r_t|$ as a random variable with uniformly 
distributed $t\geq 0$, we can invoke Markov's inequality
to conclude
\begin{eqnarray}
\Prob\big(\, 
|r_t|\leq \norm{A}\, \pmax^{1/3}
\,\big)\geq 1 - \pmax^{1/3}
\ ,
\label{14}
\end{eqnarray}
where the left hand side denotes the probability
that $|r_t|\leq \norm{A}\pmax^{1/3}$ for a
randomly drawn $t\in[0,\infty)$.
It is plausible, and has been worked out in
detail e.g. in \cite{rei12}, that essentially the same
conclusion remains true for a randomly drawn
$t\in[0,T]$, provided $T$ is sufficiently large.
Assuming 
\begin{eqnarray}
\pmax:= \max_{n\geq 1}p_n\ll 1
\ ,
\label{15}
\end{eqnarray}
it follows that the contribution
of $r_t$ to (\ref{10}) is negligible
(compared to the full range of possible values,
which $\Aobs_t$ in (\ref{9}) in principle 
could take)
for the vast majority of all sufficiently large
$t$, symbolically indicated as
$r_t \simeq 0$.
The assumption (\ref{15}) means that all energy 
levels with the possible exception of 
$|0\rangle$ must
be weakly populated, and represents,
as said in the introduction, the key
prerequisite of our present approach.

Likewise, (\ref{15}) implies for the vast majority 
of all sufficiently large $t$ that
$s_t\simeq 0$,
$\alpha_t \simeq 0$, and 
$\Aobs_t\simeq\ 
\bar\Aobs:=
\overline{\Aobs_t}$.
The latter relation is tantamount to the
equilibration results mentioned at the
beginning of the paper,
see also Refs. \cite{rei12,bal16} for its
detailed derivation.
By exploiting  (\ref{11}) and $\alpha_t \simeq 0$
in (\ref{8}),
we furthermore obtain
\begin{eqnarray}
\langle u|\psip_t\rangle \simeq 0
\, ,\ 
\langle w|\psip_t\rangle \simeq 0
\ ,
\label{16}
\end{eqnarray}
where $|u\rangle:=\sqrt{1-p_0}\,A|0\rangle$ and 
$|w\rangle:=\sqrt{1-p_0}\,A|\psi'_0\rangle$.
Introducing all these 
findings into (\ref{9}) implies
\begin{eqnarray}
\langle\psip_t|A|\psip_t\rangle  \simeq
\frac{
\bar\Aobs
- p_0 \langle 0|A|0\rangle}{1-p_0} 
\ .
\label{17}
\end{eqnarray}

Notably, the special case that {\em all\,} level 
populations $p_n$ are small is still admitted by 
our approach, and is -- in view of (\ref{15}) --
most conveniently recovered by considering
$|0\rangle$ as a purely formal ``ancillary level''
with $p_0=0$ and hence $c_0=0$ in (\ref{1}).
Moreover, all primed quantities in (\ref{2})-(\ref{7})
then coincide with their unprimed counterparts.

So far, the approximations 
(\ref{16})-(\ref{17})
only pertain to the vast majority 
of all sufficiently late times
(see below (\ref{15})).
Our next goal is to also cover 
the earlier times.

For simplicity, we first focus again on
the special case $c_0=0$ from above.
The time-evolved state $|\psi_t\rangle$ is thus 
decomposed according to (\ref{6})
into two orthonormalized components
$|\psi'_0\rangle$ and $|\psip_t\rangle$,
the first being identical and the 
second orthogonal to the initial 
state  $|\psi_0\rangle$
(see above (\ref{8})).
The main difference of the earlier times 
$t$ compared to the later ones is that
$|\psi_t\rangle$
still somehow ``remembers'' 
the specific non-equilibrium properties 
of $|\psi_0\rangle$.
But since
$|\psip_t\rangle$ is
always
orthogonal to $|\psi_0\rangle$,
it seems reasonable to expect 
that this remembrance of the 
initial state $|\psi_0\rangle$ will
mainly concern the
component of $|\psi_t\rangle$
parallel to $|\psi_0\rangle$,
while the (normalized) contribution
$|\psip_t\rangle$ orthogonal
to $|\psi_0\rangle$
will behave similarly at early 
and at later times with respect 
to some very basic properties,
such as the scalar
product with a fixed vector 
appearing in (\ref{16}), or 
the expectation value on
the left hand side of (\ref{17}).

Similar arguments
apply in the case $c_0\not=0$,
except that now there are {\em two\,} ``special 
directions'', $|\psi_0\rangle$ and $|0\rangle$, 
to which $|\psip_t\rangle$ in (\ref{6}) 
is always orthogonal.

From a different viewpoint,
the situation may also be considered as 
a natural extension
of previously established, 
non-dynamical ``typicality'' concepts 
\cite{tas16,llo88,gol06,pop06}
into the dynamical realm:
The vectors $\{|\psip_t\rangle\}_{t=0}^{\infty}$,
explore a considerable part of the
high dimensional orthogonal complement of
$|\psi_0\rangle$
and $|0\rangle$, 
hence they are typically 
(for most $t$) almost
orthogonal to a given vector
(cf. (\ref{16}))
and assume similar expectation 
values for a given
observable (cf. (\ref{17})).
For this reason, we henceforth
denote the extension of (\ref{16}) 
and (\ref{17}) to arbitrary $t$
as {\em typicality of the orthogonal
dynamics\,} (TOD).

Exploiting TOD
in (\ref{9}) 
yields -- after some straightforward but slightly
tedious algebra -- our final result
\begin{eqnarray}
&& \Aobs_t \simeq
\bar\Aobs 
+ a  f_t+a^\ast  f_t^\ast
+ | f_t|^2
 \left[
\Aobs_0\!-\!
\bar\Aobs
- a - a^\ast
\right]\,, 
\ \ \ \ \ 
\label{18}
\\
&& 
 f_t :=
\frac{e^{-i E_0t} g_t - p_0}{1-p_0}\ \, , \ \ 
 g_t := \sum_{n=0}^N p_n\, e^{i E_n t} \ ,
\label{19}
\\
&& 
a :=
c_0 \, \langle\psi_0|A|0\rangle - p_0 \, \langle 0|A|0\rangle
\ .
\label{20}
\end{eqnarray}
Specifically, if all level populations 
are small 
(see above) we thus obtain
\begin{eqnarray}
\Aobs_t \simeq \bar\Aobs + |g_t|^2
\left[\Aobs_0\!-\!\bar\Aobs \right] 
\ .
\label{21}
\end{eqnarray}

As a further corroboration of TOD, we have
rederived the same result (\ref{18})
by means of a more rigorous, but less 
enlightening and quite arduous generalization 
of the approach from \cite{rei16,bal17,rei19}:
The first step consist in skillfully 
``rearranging'' the 
large number of 
energies $\{E_n\}_{n=1}^N$
in (\ref{1}) and 
then to ``redistribute''
the corresponding 
level populations $\{p_n\}_{n=1}^N$,
yielding a very accurate approximation
in terms of an effective (auxiliary) model
with nearly equally populated eigenstates 
\cite{rei19}.
In a second step, one can show that the 
so obtained effective relaxation
dynamics is approximately invariant under
the vast majority of all permutations of 
its auxiliary eigenstates,
as worked out by way of generalizing
\cite{rei16,bal17}
in the PhD thesis \cite{bal18}.
Taking for granted that the ``true''
(non-permuted) model belongs to
that vast majority, one
eventually
recovers (\ref{18}).

Yet another confirmation of TOD
can be obtained in the small $t$ regime:
Considering the set of all
initial states $|\psi_0\rangle$ with 
the same values of $c_0$ and
$\langle\psi_0|A|\psi_0\rangle$
as the ``true'' $|\psi_0\rangle$, 
it can be shown along the lines of
\cite{bar09} that most of them 
satisfy (\ref{16}) and (\ref{17})
extremely well for $t\to 0$
under these sufficient
(but not necessary) conditions:
Among the levels $\{ |n\rangle\}_{n=1}^N$,
only those with
energies $E_n$ in some small 
(microcanonical) energy window 
are non-negligibly populated, 
and the concomitant 
diagonal and off-diagonal matrix 
elements $\langle m|A|n\rangle$ satisfy 
the eigenstate thermalization hypothesis 
(ETH) \cite{dal16,sre96}.

As a final validation of TOD,
we will compare in the next 
Sec. \ref{s4} our main prediction 
(\ref{18}) with numerical results
for a variety of specific examples.
Before doing so, it is instructive
to discuss the functions $f_t$ and 
$g_t$ from (\ref{19}) in somewhat 
more detail:

First, we can conclude
from (\ref{19}) that $f_0=1$ and 
-- similarly as in (\ref{12})-(\ref{15}) --
that $f_t\simeq 0$ for (most) 
sufficiently large $t$.
The intermediate $t$'s
thus govern the non-trivial
part of the temporal 
relaxation in (\ref{18}). 

Second, it is often useful 
to rewrite $g_t$ from (\ref{19})
by means of (\ref{1}) as
the survival amplitude of the 
initial state,
\begin{eqnarray}
g_t=\langle\psi_t|\psi_0\rangle
\ .
\label{22}
\end{eqnarray}
Since $|\psi_t\rangle$ can be obtained by 
time-evolution methods \cite{pae19,wei06},
diagonalizing the Hamiltonian is 
thus {\em not\,} mandatory to 
determine $g_t$.

Alternatively, (\ref{19}) may also be
written as
\begin{eqnarray}
g_t=\!\int \! {\rm d}E\, \rho(E)\, e^{iEt}
\, , \, \
\rho(E):=\sum_{n=0}^N p_n\,\delta(E_n-E)
\ , \ 
\label{23}
\end{eqnarray}
i.e., $\rho(E)$ describes the system's 
energy distribution and $g_t$ is its 
Fourier transform.
Note that the energy distribution is
conserved under the dynamics, hence
it can be inferred directly from the 
initial state.
Moreover, for the small-to-modelate
times $t$ during which the nontrivial
part of the relaxation takes place,
it is not mandatory to know
this distribution in all its details.
Rather, already a reasonably good 
approximation of the main features of
$\rho(E)$ will admit quite
decent predictions for the Fourier 
transform $g_t$ in (\ref{23}),
see also \cite{rei16} for 
various specific examples along these lines.

\section{Comparison with numerical examples}
\label{s4}
In the following subsections we compare our
above obtained analytical predictions within
numerical results. Moreover, we exemplify in
Sec. \ref{s43} the extension of those 
predictions to situations where more than 
one level exhibits a non-small population.

\subsection{XXZ model}
\label{s41}
Our first example is the integrable
spin-1/2 XXZ-chain with 
anisotropy parameter $\Delta$,
magnetic field $B$, 
and 
periodic boundary conditions,
\begin{eqnarray}
H=\sum_{i=1}^L
S_i^xS_{i+1}^x
+
S_i^yS_{i+1}^y+
\Delta S_i^zS_{i+1}^z
-B\,S_i^z
\ ,
\label{24}
\end{eqnarray}
where $S_i^{x,y,z}$ are the
spin operators
at lattice site
$i$. 

Focusing on $B\gg 1+\Delta$
and introducing the abbreviations 
$B':=B-\Delta/2$, $B'':=B-\Delta$,
one finds 
$|0\rangle=| \!\! \uparrow\!...\!\uparrow\rangle$
(all spins ``up'')
as ground state with energy $E_0=-B'L/2$,
followed by a first ``band'' of excited states 
$|n\rangle=\sum_{k=1}^L e_{n\cdot k}\,|s_k\rangle$
with 
energies 
$E_n=E_0+\cos(2\pi n/L)+B''$,
where $n=1,...,L$, $e_m:=e^{i 2\pi m/L}/\sqrt{L}$,
and where $|s_k\rangle$ represents the state
with all but the $k$-th spin ``up''.
Thus, the ``gap'' between this energy band 
and $E_0$ equals $B''-1\gg 1$ for even 
$L$, and is slightly larger than
$B''-1$ for odd $L$.
After another gap of comparable size,
there follows a second band, and so on.

As initial condition $|\psi_0\rangle$ we 
choose the ground state of the modified 
Hamiltonian
\begin{eqnarray}
\tilde H:=H+\Pi\, b_x S_1^x\Pi
\ .
\label{25}
\end{eqnarray}
Here, $b_x S_1^x$ models an
``impurity'' in terms of a magnetic field
$b_x$, which is perpendicular to the 
$B$ field in (\ref{24}), and
which only acts on one site of the 
spin chain.
Due to the periodic boundary conditions,
we have chosen the first site without
loss of generality.
Furthermore,
$\Pi:=\sum_{n=0}^L|n\rangle\langle n|$
in (\ref{24}) 
is the projector on the subspace spanned by 
the ground state and the first band of excited 
states of $H$,
and therefore eliminates the effects of 
the second  and higher bands.
Focusing on $B''\gg 1$ and
not too large $b_x$, it is
reasonable to expect, and can be confirmed 
by a more detailed calculation, that
these omitted effects are indeed 
very weak.
Altogether, our setup thus still amounts to
a physically sensible quantum quench 
scenario (see introduction)
with post-quench Hamiltonian (\ref{21}),
pre-quench Hamiltonian (\ref{25})
and exhibiting one non-small level 
population.

Given this setup, a straightforward calculation 
then yields (without any further approximation)
$|\psi_0\rangle=\nu\sum_{n=0}^L\gamma_n|n\rangle$,
where $\nu:=(\sum_{n=0}^L|\gamma_n|^2)^{-1/2}$,
$\gamma_0:=1$,
$\gamma_{n\geq 1}:=-\sqrt{\eta/L}\,e^\ast_n\,[E_n-E_0+\eta q]^{-1}$,
$\eta:=(b_x/2)^2$, and where $q$ is the smallest
positive solution of the transcendental equation
$Lq=\sum_{n=1}^L[E_n-E_0+\eta q]^{-1}$.
Solving this equation, and finally evaluating
the exact expectation values according to
Eqs. (\ref{1}) and (\ref{9}), and their approximative
counterparts according to (\ref{18})-(\ref{20}) 
is an easy numerical task up to rather 
large $L$ values.

\begin{figure}[tb]
\epsfxsize=1.0\columnwidth
\epsfbox{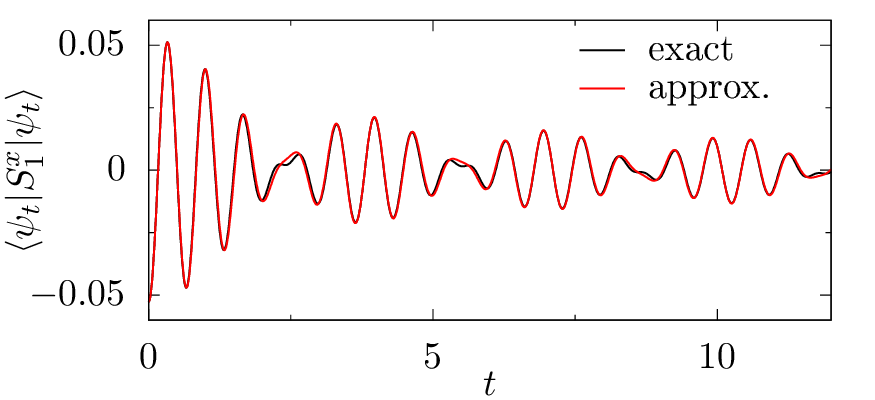}
\caption{\label{fig1}
Time-dependent expectation values of 
the observable $A=S_1^x$
according to the exact Eqs. (\ref{1}), (\ref{9})
and the approximation (\ref{18})
for the XXZ chain from (\ref{24}) with 
$L=1000$, $B=10$, and $\Delta=0.5$.
The initial state $|\psi_0\rangle$ 
was generated by a quantum quench 
as detailed in the main text,
resulting in $p_0\simeq 0.997$ 
and $\pmax\simeq 3.5\cdot 10^{-6}$.
}
\end{figure}

The so obtained results 
for the numerically exact
expectation values of $A=S_1^x$
together with the corresponding
analytical approximation from 
(\ref{18})
are exemplified in Fig. \ref{fig1}
as black and red lines, respectively.
The main observations are that the
temporal decay exhibits a quite rich
structure and that the analytics
captures them remarkably well.

For the specific example in Fig. \ref{fig1},
we chose $b_x=1$ in (\ref{24}),
and the numerically obtained quantitative 
values for the ground state population
$p_0$ and for the second largest 
population $\pmax$ from (\ref{15}) are
$p_0\simeq 0.997$ and 
$\pmax\simeq 3.5\cdot 10^{-6}$,
respectively.
Moreover, closer inspection 
of the numerical data 
(beyond the range displayed in 
Fig. \ref{fig1})
reveals that the oscillations of
$\langle\psi_t|A|\psi_t\rangle$
decrease rather slowly.
Quantitatively, the envelopes
seem to asymptotically 
approach
$\bar\Aobs=0$ 
in Fig. \ref{fig1} essentially as 
$t^{-1/2}$ for large times.
We finally mention that further increasing 
$L$ in Fig. \ref{fig1} did not result in 
any notable changes of the two curves.

\subsection{Modified XXZ model}
\label{s42}
As a second example, we omit the $B$-field in (\ref{24})
and include next-nearest neighbor interactions, yielding
\begin{eqnarray}
H=\sum_{i=1}^L
S_i^xS_{i+1}^x
+
S_i^yS_{i+1}^y
+
\Delta S_i^zS_{i+1}^z
+
\Delta' S_i^zS_{i+2}^z
\label{26}
\end{eqnarray}
with periodic boundary conditions,
and spin-1/2 operators $S_i^{x,y,z}$
acting on lattice site $i$.

This model is integrable for $\Delta'=0$ 
and non-integrable otherwise.
Given $\Delta'$, the energy spectrum exhibits
a gap for sufficiently large $\Delta$
(for instance $\Delta > 1$ when $\Delta'=0$
\cite{maz16}).
More precisely, the two lowest energies are
almost degenerate (approaching an exact degeneracy 
for $L\to\infty)$, and are separated by a gap
(which persist for $L\to\infty)$
from all
other energies 
\cite{mikeska2004}.
For the large but finite $L$'s
and small-to-moderate $t$'s of interest
to us, the two almost degenerate lowest energies can be
safely approximated as being strictly degenerate
(a rigorous bound for the corrections
is provided in Appendix B of \cite{rei19}).
As detailed below (\ref{1}), the situation 
thus effectively amounts to a single (non-degenerate)
ground state, separated by a gap from the excited
states.
As initial condition $|\psi_0\rangle$ we choose the
N\'eel state 
$|\!\!\uparrow\downarrow\uparrow\!...\!\downarrow\rangle$
(tensor product of alternating single-spins
``up'' and ``down''), 
and we tacitly focus on even $L$ from now on.
At the same time, this $|\psi_0\rangle$ 
is the ground state of (\ref{26}) when 
$\Delta \to \infty$,
i.e., our setup may again be viewed 
as a quantum quench scenario
and exhibits a non-small population $p_0$ 
for sufficiently large post-quench 
values of $\Delta$ in (\ref{26})
\cite{maz16, fagotti2014, barmettler2009, piroli2017, pozsgay2014, wouters2014}.

\begin{figure}[tb]
\epsfxsize=1.0\columnwidth
\epsfbox{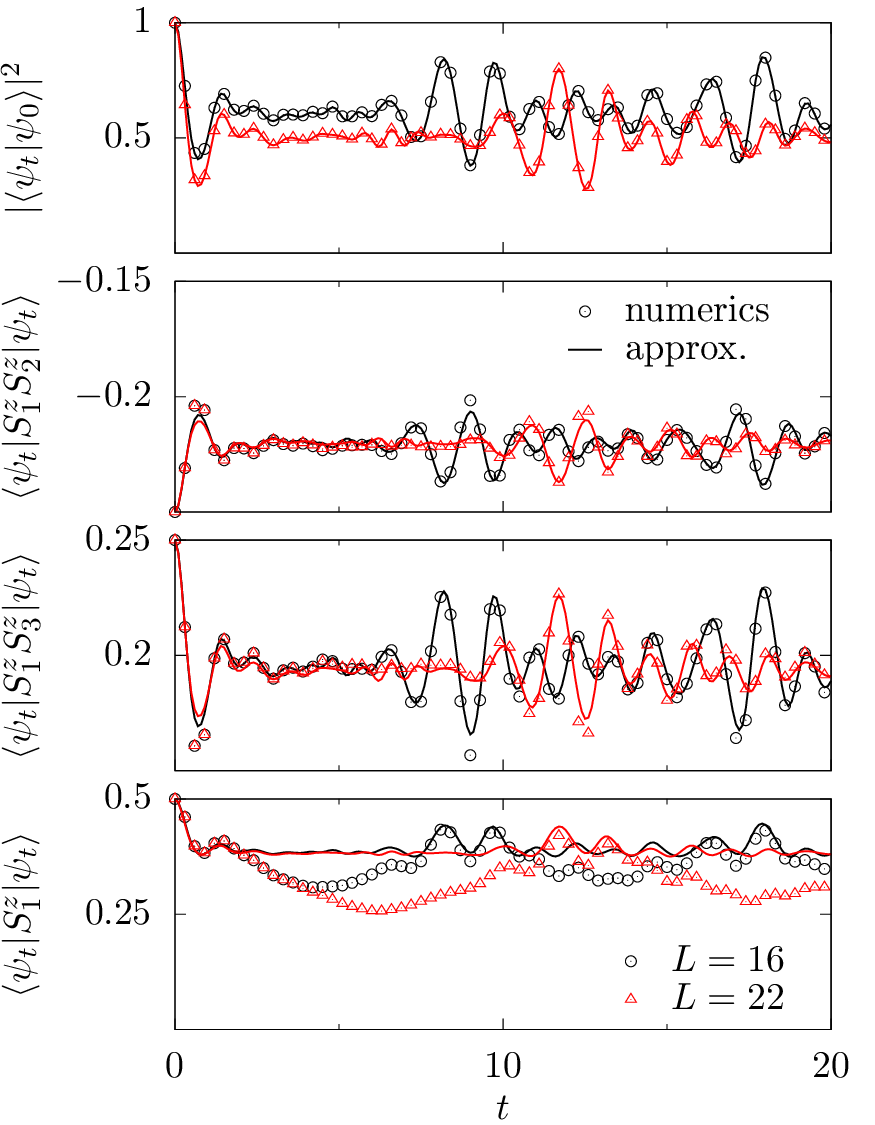}
\caption{\label{fig2}
Time-dependent expectation values of
the observables
$A=|\psi_0\rangle\langle\psi_0|$,
$A=S^z_1S^z_2$,
$A=S^z_1S^z_3$, and
$A=S^z_1$
for the non-integrable 
spin-chain model from 
(\ref{26}) with $\Delta =3$,
$\Delta'=-0.5$,
$L=16$ (black), $L=22$ (red),
and initial condition
\mbox{$|\psi_0\rangle=|\!\!\uparrow\downarrow\uparrow\!...\!\downarrow\rangle$}
(N\'eel state).
Symbols: Numerical results
as detailed in 
the main text.
Lines: Analytical approximation
(\ref{18}).
}
\end{figure}

\begin{figure}[tb]
\epsfxsize=1.0\columnwidth
\epsfbox{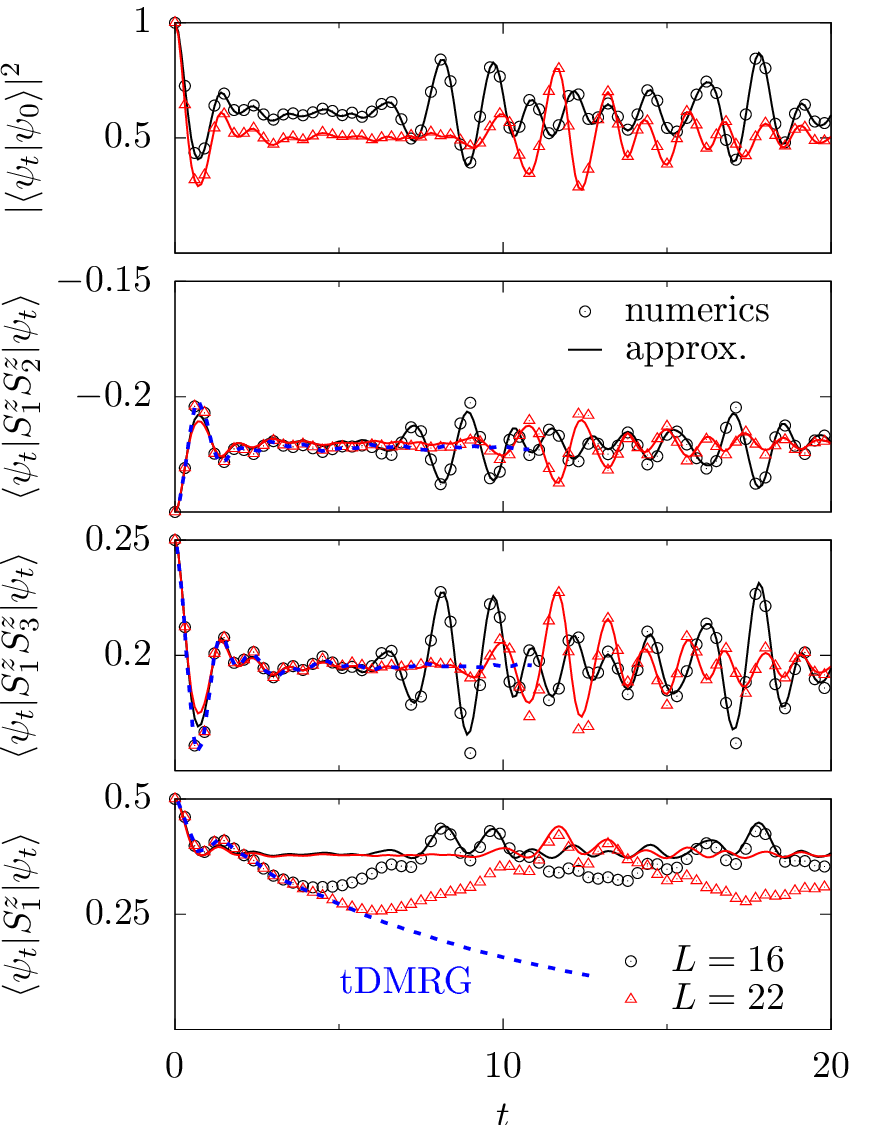}
\caption{\label{fig3}
Same as Fig. \ref{fig2} but
for an integrable model 
with $\Delta =4$ 
and $\Delta'=0$.
Moreover, the time-dependent density matrix 
renormalization group (tDMRG) results
from Refs.~\cite{fagotti2014, barmettler2009} 
are indicated
as dashed blue lines.
}
\end{figure}

Employing numerical exact diagonalization (ED) 
of the Hamiltonian \eqref{26} for moderate 
system sizes $L$, it is straightforward to simulate 
the time evolution of the initial state 
$|\psi_0\rangle$ (cf. Eq. (\ref{1})), 
and to compute the expectation values of the 
desired operators (cf. Eq. (\ref{9})). 
Moreover, since ED yields the ground state 
$|0\rangle$, as well as all $p_n$ 
and $E_n$, also our theoretical prediction 
(\ref{18})-(\ref{20}) can be readily evaluated.
Both the numerical data and the analytical 
approximations for $L=16$ in
Figs. \ref{fig2} and \ref{fig3} have been
obtained along these lines.

As usual, since the $z$-component of the total
spin is conserved for all values of $\Delta$ and 
$\Delta'$ in (\ref{26}), i.e., $[H,\sum_i S_i^z] = 0$, 
we can restrict ourselves to the diagonalization 
of the zero-magnetization subsector.
Nevertheless, ED is practically feasible only 
up to relatively small (even) $L$ values.

In order to also cover larger $L$ values, 
we adopted an alternative numerical approach.
The first key point is to utilize for the function
$g_t$ in (\ref{19}) the alternative expression
in terms of the survival amplitude from 
(\ref{22}) (see also (\ref{31}) below).
Hence, full ED can be circumvented by 
solving the time-dependent Schr\"odinger 
equation iteratively, e.g., by means of a 
Runge-Kutta scheme with small time step
\cite{steinigeweg2015}, 
or also by other sophisticated approaches 
\cite{weisse2006}. 
The quantities $|0\rangle$, $c_0$, and $p_0$
in (\ref{19}) and (\ref{20}) can then be obtained 
by means of standard Krylov-subspace 
techniques. 
Both the numerical data and the analytical 
approximations for $L=22$ in
Fig. \ref{fig2} and Fig. \ref{fig3} 
have been obtained along these lines.

In Figs. \ref{fig2} and \ref{fig3} 
we compare the so obtained
numerical results with the analytical
prediction from (\ref{18}) for the four
observables $A=|\psi_0\rangle\langle\psi_0|$
(survival probability of the initial state),
$A=S^z_1S^z_2$ (correlation of neighboring spins),
$A=S^z_1S^z_3$ (next-nearest neighbor correlation),
and
$A=S^z_1$ (single spin). 
(Note that the specific lattice site is arbitrary 
due to the periodic boundary conditions 
in (\ref{26}) and the initial N\'eel state.)
Our first observation is that the 
non-integrable and integrable 
examples in Fig. \ref{fig2} and 
Fig. \ref{fig3} behave quite similarly.
Second, the agreement between
our numerical solutions 
(for $L=16$ and $L=22$)
and the theoretical approximation 
is quite good, at least for the first 
three observables 
$A=|\psi_0\rangle\langle\psi_0|$,
$A=S^z_1S^z_2$, and
$A=S^z_1S^z_3$
(the fourth observable will 
be discussed in more detail later)

As opposed to Fig. \ref{fig2}, in the case
depicted in Fig. \ref{fig3} also time-dependent
density matrix renormalization group 
(tDMRG) results are available
from the literature \cite{fagotti2014, barmettler2009} 
and are shown as dashed blue lines.
In contrast to our own numerics
for $L=16$ and $L=22$,
these tDMRG studies from 
\cite{fagotti2014, barmettler2009}
were designed to approximate 
the thermodynamic limit $L\to\infty$.

In view of those results for $L\to\infty$,
and since our main objective is to illustrate 
the validity of the theoretical prediction (\ref{18}) 
for large but fixed values of $L$, we refrained 
from a more detailed finite-size scaling 
analysis.

In all depicted cases, we numerically 
found $p_0\simeq 0.78$ for the ground 
state population 
and $\pmax\simeq 0.05$ for the second largest 
population from (\ref{15}).
More precisely, both values decrease 
with increasing $L$ but only extremely 
slowly.
On the other hand, we recall that
$\pmax$ must be sufficiently small
to guarantee equilibration
(see introduction), and to 
satisfy our present condition (\ref{15}).
The fact that $\pmax\simeq 0.05$ is still 
not really small explains why
the expectation  values in
Figs. \ref{fig2} and \ref{fig3}
still do not equilibrate too well for 
large $t$, 
but rather keep ``oscillating''
quite notably about their 
long-time average.
Moreover, this also seems to be
a main reason for the remaining 
deviations of (\ref{18})
from the numerics in 
Figs. \ref{fig2} and \ref{fig3}.
Unfortunately, substantially smaller
$\pmax$ values would require 
numerically unfeasibly large 
$L$'s.
For the rest, we still find it remarkable
that the prediction
(\ref{18}) often reproduces 
quite reasonably even those 
numerically obtained oscillations 
at large $t$.
Apparently, the requirement of
small $\pmax$ values is quantitatively
less stringent in (\ref{15}) than
with respect to equilibration.

In other words, the remnant ``oscillations'' 
for large $t$ may be interpreted as finite
$L$-effects, which, similarly as $\pmax$, 
decrease only very slowly with 
increasing $L$.
Besides our numerics, also the tDMRG results 
are in agreement with
(and thus provide further support to)
this interpretation.
More precisely speaking, our finite $L$ results
compare quite well with the tDMRG
($L\to\infty$) approximations 
for small-to-moderate 
$t$, but start do deviate for
larger $t$ values by developing
the above mentioned ``oscillations''.
With increasing $L$, the onset
of those deviations moves towards 
larger times.
The latter effect is in fact considerably
more pronounced than the simultaneous,
but much weaker reduction of the
oscillation amplitude with 
increasing $L$.

Turning to the last observable
$A=S^z_1$ in Figs. \ref{fig2} and \ref{fig3},
the agreement between
our numerical solutions 
(for $L=16$ and $L=22$)
and the theoretical approximation
is still satisfying 
for small $t$ and to some extent 
also for large $t$.
In fact, the agreement with 
an appropriately generalized 
theory becomes again very 
good for even larger times, 
as we will see in the next 
subsection.
On the other hand, for the 
intermediate times there remains a
notable disagreement between
theory and numerics, but also 
between our numerics and 
the tDMRG results.
Apparently, we are dealing with 
some quite subtle 
and obstinate finite $L$ effects for 
moderate-to-large $t$ 
(see also Ref.\ \cite{barmettler2009}).
More precisely speaking,
the reason for 
those deviations seems to be that in 
this specific example the limits 
$t\to\infty$ and
$L\to\infty$ apparently do not commute.
Indeed, focusing first on the thermodynamic 
limit $L\to\infty$, it seems
quite reasonable to expect, and is 
also supported by the tDMRG results in
Fig.~\ref{fig3}, that the expectation value
$\langle \psi_t|S^z_1|\psi_t\rangle$
approaches zero for large times $t$.
On the other hand, the numerical 
results for finite $L$ in Fig.~\ref{fig3} 
substantially deviate 
from the latter relaxation behavior
of the tDMRG data
beyond some ``critical'' time, which
increases with $L$.
(In fact, this crossover seems to 
be closely connected with the onset of 
the above mentioned ``oscillations''.)
Therefore, it may not be so surprising 
that our simple theory misses those quite
subtle effects of the competition 
between the two non-commuting 
limits, especially in the transition
region (moderate $t$ in Figs. \ref{fig2} and \ref{fig3})
between the regime where the large $L$
limit ``wins'', and the regime where
the large $t$ behavior at finite $L$ 
takes over.

\subsection{Two levels with non-small populations}
\label{s43}
We consider the same setup (and notation)
as before,  except that now 
{\em two} levels, namely those with indices 
$n=0$ and $n=1$, may exhibit non-small 
populations $p_0$ and $p_1$.
As explained below Eq. (\ref{1}), we again
can and will exclude degeneracies. 
In particular, we assume that $E_0\not=E_1$.

In principle, the extension of the TOD concept 
from Sec. \ref{s3} is straightforward, however, 
carrying out the actual calculations is a quite
arduous task (see also above Eq. (\ref{18})). 
Omitting them here, one finally obtains
\begin{eqnarray}
\Aobs_t & \simeq &
\bar\Aobs + \kappa_t+\kappa^\ast_t
+ |\tilde f_t|^2 
\left[
\Aobs_0\!-\!\bar\Aobs\!-\!\kappa_0\!-\!\kappa_0^\ast
\right]\, ,
\ \ \ \
\label{29}
\\[0.1cm]
\tilde f_t & := & 
\frac{g_t - p_0\, e^{iE_0t}
- p_1\, e^{iE_1t}}{1-p_0-p_1}
\, ,
\label{30}
\\[-0.1cm]
g_t  & := & 
\sum_{n=0}^N p_n\, e^{i E_n t} 
=\langle\psi_t|\psi_0\rangle
\, ,
\label{31}
\\
\kappa_t & := & 
\lambda_0\chi_0(t) + \lambda_1\chi_1(t) + \gamma\, e^{i(E_1-E_0)t}
\, ,
\label{32}
\\[0.1cm]
\lambda_0 & := & 
c_0 \, \langle\psi(0)|A|0\rangle - p_0 \, \langle 0|A|0\rangle - \gamma
\, ,
\label{35}
\\[0.1cm]
\lambda_1 & := & 
c_1 \, \langle\psi(0)|A|1\rangle - p_1 \, \langle 1|A|1\rangle - \gamma^\ast
\, ,
\label{36}
\\[0.1cm]
\gamma & := & c_0c_1^\ast  \langle 1|A|0\rangle
\, ,
\label{37}
\\[0.1cm]
\chi_{\nu}(t) & := & 
\tilde f_t\, e^{-iE_{\nu}t}\ \ (\nu=0,1)
\, .
\label{34}
\end{eqnarray}

In analogy to Eq. (\ref{15}), the main precondition 
for the above results now takes the form
\begin{eqnarray}
\max_{n\geq 2}p_n\ll 1
\label{39}
\end{eqnarray} 

By means of a similar arguments as 
between Eq. (\ref{15}) and Eq. (\ref{16}),
one can infer from (\ref{30}) that 
$\tilde f_t\simeq 0$ for the vast majority 
of all sufficiently large $t$.
The same property is inherited by 
$\chi_{\nu}(t)$ in (\ref{34}), hence
$\kappa_t$ in (\ref{32}) is dominated 
by the last term on the right 
hand side. Finally, (\ref{29}) can then
be approximated for most sufficiently 
large $t$ as
\begin{eqnarray}
\Aobs_t & \simeq & \bar\Aobs + 
\gamma\, e^{i\omega t} + \gamma^\ast e^{-i\omega t}
\label{46}
\\
\omega & := & (E_1-E_0)/\hbar
\ ,
\label{47}
\end{eqnarray}
i.e., $\Aobs_t$ oscillates about the temporal
mean value $\bar\Aobs$ with frequency $\omega$ 
and amplitude $2|\gamma|$ (see also (\ref{37})).

The previously obtained results 
for a single level with non-small 
population $p_0$ are readily recovered 
by formally setting $p_1=0$ and 
thus $c_1=0$ (see also discussion 
below Eq. (\ref{17})).

\begin{figure}[tb]
\epsfxsize=1.0\columnwidth
\epsfbox{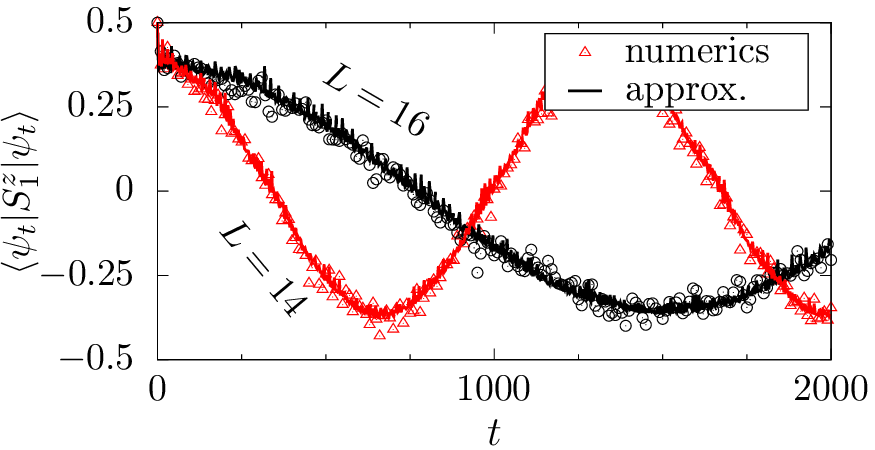}
\caption{\label{FigS3} 
Time-dependent expectation values of $A = 
S_1^z$ for $\Delta = 4, \Delta' = 0$ (analogous 
to bottom panel of Fig.\ 
\ref{fig3}, but now for times up to $t=2000$),
and for system sizes $L = 14$ and $L=16$. 
The numerically observed slow oscillations of 
$\langle \psi_t|A|\psi_t\rangle$ (resulting 
from the level splitting between the two 
almost-degenerate eigenstates) are remarkably 
well captured by our generalized analytical
approximation from (\ref{29})-(\ref{34}).}
\end{figure}

As an example we consider again the 
model from (\ref{26}) with a N\'eel state 
$|\!\!\uparrow\downarrow\uparrow\!...\!\downarrow\rangle$
as initial condition.
Focusing again on the gapped regime,
the two lowest energies of this model
are for large $L$ almost degenerate, 
approaching an exact degeneracy for 
$L\to\infty$, and are (almost)
equally populated by the N\'eel state.
For the reasonably large $L$'s
and small-to-moderate $t$'s which we 
considered so far, those two almost 
degenerate lowest energies could 
thus always be safely approximated 
as being strictly degenerate.
Indeed, our generalized theoretical 
prediction
(\ref{29})-(\ref{34}) would be practically
indistinguishable from the solid lines
in Figs. \ref{fig2} and \ref{fig3}.

Therefore, we now turn to even much 
larger times $t$, for which the difference
between the two lowest energies is no 
longer negligible, and hence the 
generalized theory (\ref{29})-(\ref{36}) 
must be employed.
In doing so, we adopted the same numerical 
ED methods as described in detail in the
previous subsection.

Fig.\ \ref{FigS3} shows the so obtained
time-dependent expectation values
$\langle \psi_t|S^z_1|\psi_t\rangle$
for two 
different system sizes $L = 14$ and $L=16$.
In particular, the black line and symbols 
($L=16$) correspond to those in the bottom 
panel of Fig.\ \ref{fig3}, except that they 
now cover a much larger time interval
up to $t=2000$.
Most notably, for such long times, 
we find that the observable
$A=S^z_1$ gives rise to approximately 
harmonic oscillations,
in good agreement with the
theoretical asymptotics
(\ref{46}).
In particular, the oscillations
persist up to arbitrarily large times,
i.e., we are dealing with an example
which does not exhibit equilibration
in the long-time limit.

As said above, the energy difference 
between the almost degenerate levels
is known to decrease with increasing $L$.
In agreement with (\ref{47}), 
the oscillation frequency 
in Fig.\ \ref{FigS3} is indeed
observed to decrease with 
increasing~$L$.

Finally, and again in close analogy
to the previous subsection, we numerically
found that $p_0\simeq p_1\simeq 0.39$, 
$\max_{n\geq 2}p_n\simeq 0.05$
and that both values decrease 
very slowly with $L$.
Accordingly, the amplitude of the
oscillations in Fig. \ref{FigS3}
decreases very little with 
increasing~$L$.

\subsection{Small populations of all levels}
\label{s44}
In this subsection, we further elaborate on the
special case that {\em all} level populations 
$p_n$ are small (see also below Eq. (\ref{17})),
including a comparison of numerical results
with the corresponding simplified analytical
approximation from (\ref{21}).


\begin{figure}
\epsfxsize=0.95\columnwidth
\epsfbox{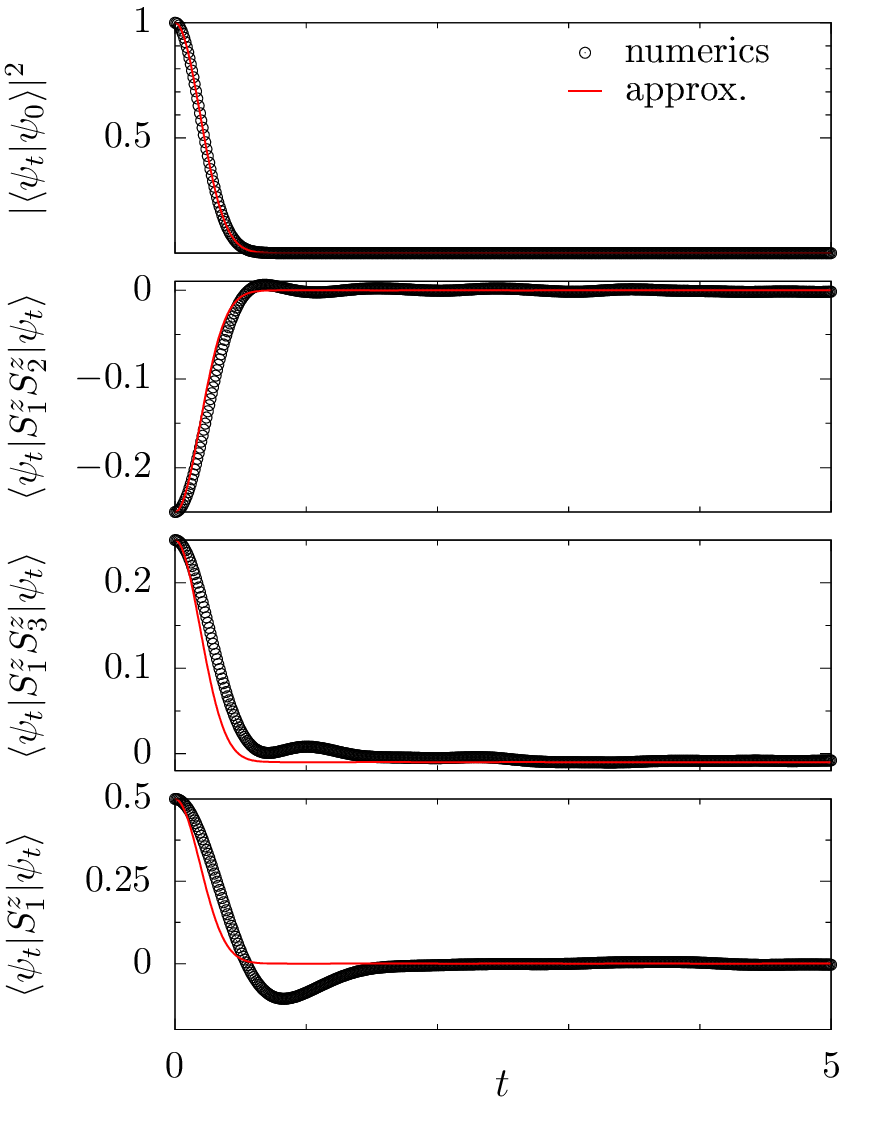}
\caption{\label{FigS4} 
Time-dependent expectation values 
of $A = |\psi_0\rangle \langle \psi_0|$, 
$A = S_1^x S_2^x$, 
$A = S_1^z S_3^z$, and 
$A = S_1^z$ for the fully-connected model 
\eqref{Eq::XYZ} with $L = 16$ sites and a 
fixed choice of the random couplings
$J_{ij}^{x,y,z}$.
Symbols: Numerical results (ED).
Lines: Analytical approximation (\ref{21}).}
\end{figure}

To this end, we consider the fully connected 
spin model
\begin{equation}\label{Eq::XYZ}
 H = \sum_{i = 1}^L \sum_{j > i}^L J_{ij}^x S_i^x S_j^x + J_{ij}^y S_i^y S_j^y 
+ J_{ij}^z S_i^z S_j^z\ ,
\end{equation}
where the couplings $J_{ij}^{x,y,z}$ are independent,
Gaussian distributed random variables with zero mean 
and unit variance,
and we choose again a N\'eel state 
$|\!\!\uparrow\downarrow\uparrow\!...\!\downarrow\rangle$
as initial condition.
Two important differences in comparison
with the previous model (\ref{26}) are:
First, the random couplings eliminate any 
spatial structure and lift all symmetries of 
the model so that the initial N\'eel state 
is randomly spread over the whole energy 
basis with all $p_n$ being very small 
quantities on the order of $2^{-L}$.
Second, the total magnetization is generally 
not any more conserved ($[\sum_i S_i^z, H] \neq 0$)
so that the resulting dynamics has to be understood 
with respect to the full Hilbert space with 
dimension $2^L$. 
Apart from that, we employed the same 
numerical ED methods as before.

For this model and initial condition, we depict in 
Fig.~\ref{FigS4} our numerical and
analytical results for the 
usual four observables 
$A=|\psi_0\rangle\langle\psi_0|$,
$A=S^z_1S^z_2$, 
$A=S^z_1S^z_3$, 
and $A=S^z_1$, where we have 
used one fixed realization of 
the random couplings $J_{ij}^{x,y,z}$
in (\ref{Eq::XYZ}).
Moreover, since finite-size effects are 
expected to be relatively weak in such 
a random and fully-connected model, we 
restrict our analysis to a single 
system size $L = 16$. 
For all observables shown in Fig.~\ref{FigS4}, 
we find that $\langle \psi_t|A|\psi_t\rangle$ 
exhibits a fast decay on short times 
scales $t \lesssim 1$, before equilibrating 
to its long-time value $\bar\Aobs \approx 0$. 
Our analytical approximation explains the 
numerical findings in Fig. \ref{FigS4}
comparably well as in the previous cases 
in Figs. \ref{fig2} and \ref{fig3}.

The example from Fig.~\ref{FigS4} thus 
illustrates that our 
theory also covers cases where all 
$p_n$ are small, i.e., beyond the situations 
at the actual focus of our paper,
where one energy level exhibits 
a non-small population.
For a more detailed exploration of
the conceptual premises of such a 
theoretical approach we also refer 
to \cite{rei19}.
In particular, it turns out that both the 
considered observable and the initial 
state must be ``sufficiently far'' from 
any conserved quantity, since any such
too close constant of motion 
would give rise to a slower 
temporal relaxation than predicted 
by our present theory.
In the same vein, the special
``regularity'' of the initial 
N\'eel state may well be the cause
of the remaining deviations between
numerics and theory in 
Fig.~\ref{FigS4}.

\section{Conclusions}
\label{s5}
While thermal 
equilibrium properties of gapped
systems at low energies 
or in the ground state
have been previously explored 
in considerable detail,
the focus of our present work is
on the temporal relaxation of a
far from equilibrium initial 
state.
We put forward a new kind of 
typicality principle, named typicality 
of the orthogonal dynamics (TOD),
which governs
the equilibration of
isolated many-body quantum systems:
The (normalized) component of 
the system state $|\psi_t\rangle$ 
orthogonal to the initial state 
$|\psi_0\rangle$,
and possibly also to one 
non-negligibly populated 
energy level $|0\rangle$, 
typically exhibits
similar properties 
at early and at later times.
As a consequence of TOD we obtained 
an analytical prediction for the 
temporal relaxation behavior,
comparing very favorably with
a variety of numerical test cases.
Particular emphasis was laid on the
previously hardly explored relaxation of
systems with a non-small ground state 
population due to a gap in the 
energy spectrum.
However, our prediction also covers
considerably more general situations.
In particular, the system may or may 
not be integrable and thus may or 
may not exhibit thermalization.
In either case, a key role is
played by the overlaps in (\ref{22}) 
and/or the energy distribution 
in (\ref{23}).
Finally, we exemplified in 
Sec. \ref{s43} that our 
present approach can also be 
extended to situations where
several energy levels
exhibit non-small populations
and hence the system is not 
even expected to equilibrate.

\begin{acknowledgments}
We are indebted to Masud Haque for 
additional explanations and numerical 
results concerning Ref. \cite{maz16},
J\"urgen Schnack for bringing
Ref. \cite{kri14} to our attention, and
Thomas Dahm for illuminating discussions.
This work was supported by the 
Deutsche Forschungsgemeinschaft (DFG)
within the Research Unit FOR 2692
under Grants No. 397303734,
397067869, and 355031190.
\end{acknowledgments}


\begin{thebibliography}{99}

\bibitem{gog16}
C. Gogolin and J. Eisert,
Equilibration, thermalization, and the emergence
of statistical mechanics in closed quantum systems,
Rep. Prog. Phys. {\bf 79}, 056001 (2016).

\bibitem{dal16}
L. D'Alessio, Y. Kafri, A. Polkovnikov, and M. Rigol,
From Quantum Chaos and Eigenstate Thermalization
to Statistical Mechanics and Thermodynamics,
Adv. Phys.  {\bf 65}, 239 (2016).

\bibitem{tas16}
H. Tasaki,
Typicality of Thermal Equilibrium and
Thermalization in Isolated Macroscopic 
Quantum Systems,
J. Stat. Phys. {\bf 163}, 937 (2016).

\bibitem{bor16}
F. Borgonovi, F. M. Izrailev, L. F. Santos, and V. G. Zelevinsky,
Quantum chaos and thermalization in isolated systems of 
interacting particles,
Phys. Rep.  {\bf 626}, 1 (2016).

\bibitem{lan16}
T. Langen, T. Gasenzer, and J. Schmiedmayer,
Prethermalization and universal dynamics in 
near-integrable quantum systems,
J. Stat. Mech. 064009 (2016).

\bibitem{mor18}
T. Mori, T. N. Ikeda, E. Kaminishi, and M. Ueda,
Thermalization and prethermalization 
in isolated quantum systems: a theoretical overview,
J. Phys. B  {\bf 51}, 112001 (2018).

\bibitem{hob71}
A. Hobson,
{\em Concepts in Statistical Mechanics}
(Gordon and Breach, New York, 1971).

\bibitem{rei08}
P. Reimann, 
Foundation of statistical mechanics under 
experimentally realistic conditions,
Phys. Rev. Lett. {\bf 101}, 190403 (2008).

\bibitem{lin09}
N. Linden, S. Popescu, A. J. Short, and A. Winter, 
Quantum mechanical evolution towards equilibrium.
Phys. Rev.  E {\bf 79}, 061103 (2009).

\bibitem{sho11}
A. J. Short, 
Equilibration of quantum systems and subsystems, 
New J. Phys. {\bf 13}, 053009 (2011).

\bibitem{sho12}
A. J. Short and T. C. Farrelly,	
Quantum equilibration in finite time, 
New J. Phys. {\bf 14}, 013063 (2012).

\bibitem{rei12}
P. Reimann and M. Kastner, 
Equilibration of macroscopic quantum systems, 
New J. Phys. {\bf 14}, 043020 (2012).

\bibitem{bal16}
B. N. Balz and P. Reimann, 
Equilibration of isolated many-body quantum systems with 
respect to general distinguishability measures, 
Phys. Rev. E {\bf 93}, 062107 (2016).

\bibitem{maz16}
P. P. Mazza, J.-M. St\'ephan, E. Canovi, V. Alba, M. Brockmann, 
and M. Haque,
Overlap distributions for quantum quenches in the anisotropic Heisenberg chain,
J. Stat. Mech. P013104 (2016).

\bibitem{kri14}
V. Ya. Krivnov, D. V. Dmitriev, S. Nishimoto, 
S.-L. Drechsler, and J. Richter,
Delta chain with ferromagnetic and antiferromagnetic 
interactions at the critical point,
Phys. Rev. B {\bf 90}, 014441 (2014).

\bibitem{f1}
There may be isolated $t$'s with
$\beta'_t=0$ (for instance $t=0$),
for which $|\psip_t\rangle$  
is defined via continuation.

\bibitem{llo88}
S. Lloyd,
Pure state quantum statistical mechanics and black holes,
Ph.D. Thesis, The Rockefeller University, 1988,
Chapter 3, arXiv:1307.0378.

\bibitem{gol06}
S. Goldstein, J. L. Lebowitz, R. Tumulka, and N. Zangh\`{\i},
Canonical typicality,
Phys. Rev. Lett.  {\bf 96},  050403 (2006).

\bibitem{pop06}
S. Popescu, A. J. Short, and A. Winter,
Entanglement and the foundations of statistical mechanics,
Nat. Phys.  {\bf 2},  754 (2006).

\bibitem{rei16}
P. Reimann, 
Typical fast thermalization processes in closed many-body systems,
Nat. Commun. {\bf 7}, 10821 (2016).

\bibitem{bal17}
B. N. Balz and P. Reimann,
Typical relaxation of isolated many-body systems which do not thermalize,
Phys. Rev. Lett. {\bf 118}, 190601 (2017).

\bibitem{rei19}
P. Reimann,
Transportless equilibration in isolated many-body quantum systems, 
New J. Phys. {\bf 21}, 053014 (2019).

\bibitem{bal18}
B. N. Balz,
Dynamik von Quanten-Vielteilchensystemen;
Equilibration, Thermalisierung und Typikalit\"at in 
quanten-statistischer Mechanik,
Ph.D Thesis, Universit\"at Bielefeld 2018,
pub.uni-bielefeld.de/record/2930989.

\bibitem{bar09}
C. Bartsch and J. Gemmer, 
Dynamical Typicality of Quantum Expectation Values, 
Phys. Rev. Lett. {\bf 102}, 110403 (2009).

\bibitem{sre96}
M. Srednicki, 
Thermal fluctuations in quantized chaotic systems, 
J. Phys. A {\bf 29}, L75 (1996).

\bibitem{pae19}
S. Paeckel, T. K\"ohler, A. Swoboda, 
S. R. Manmana, U.~Schollw\"ock, and C. Hubig,
Time-evolution methods for matrix-product states,
Ann. Phys. {\bf 411}, 167998 (2019).

\bibitem{wei06}
A. Wei\ss e, G. Wellein, A. Alvermann, and H. Fehske,
The kernel polynomial method, 
Rev. Mod. Phys. {\bf 78}, 275 (2006).

\bibitem{mikeska2004}
H.-J. Mikeska and A. K. Kolezhuk, 
One-dimensional magnetism, in: 
Quantum Magnetism (Springer, Berlin, Heidelberg, 2004).

\bibitem{fagotti2014}
M. Fagotti, M. Collura, F. H. L. Essler, and P. Calabrese, 
Relaxation after quantum quenches in the 
spin-$\tfrac{1}{2}$ Heisenberg XXZ chain, 
Phys. Rev. B {\bf 89}, 125101 (2014). 

\bibitem{barmettler2009}
P. Barmettler, M. Punk, V. Gritsev, E. Demler, and E. Altman, 
Quantum quenches in the anisotropic spin-1/2 Heisenberg chain: different approaches to 
many-body dynamics far from equilibrium,
New. J. Phys.  {\bf 12}, 055017 (2010).

\bibitem{piroli2017}
L. Piroli, E. Vernier, P. Calabrese, and M. Rigol, 
Correlations and diagonal entropy after quantum quenches in XXZ chains, 
Phys. Rev. B {\bf 95}, 054308 (2017).

\bibitem{pozsgay2014}
B. Pozsgay, M. Mesty\'an, M. A. Werner, M. Kormos, G. Zar\'and, and G. 
Tak\'acs, 
Correlations after Quantum Quenches in the XXZ Spin Chain: Failure of the
Generalized Gibbs Ensemble, 
Phys. Rev. Lett. {\bf 113}, 117203 (2014).

\bibitem{wouters2014}
B. Wouters, J. De Nardis, M. Brockmann, D. Fioretto, M. Rigol, and J.-S. 
Caux, 
Quenching the Anisotropic Heisenberg Chain: Exact Solution and Generalized Gibbs 
Ensemble Predictions, 
Phys. Rev. Lett. {\bf 113}, 117202 (2014).

\bibitem{steinigeweg2015}
R. Steinigeweg, J. Gemmer, and W. Brenig, 
Spin and energy currents in integrable and nonintegrable 
spin-$\tfrac{1}{2}$ chains: A typicality approach to real-time 
autocorrelations, 
Phys. Rev. B {\bf 91}, 104404 (2015).

\bibitem{weisse2006}
A. Wei\ss e, G. Wellein, A. Alvermann, and H. Fehske,
The kernel polynomial method, 
Rev. Mod. Phys. {\bf 78}, 275 (2006).

\end{thebibliography}
\end{document}